\documentclass[aip, twocolumn, amsmath, amssymb, floatfix, superscriptaddress, reprint]{revtex4-1}
\usepackage{graphicx}
\begin{document}
\title{A Near-Infrared 64-pixel Superconducting Nanowire Single Photon Detector Array with Integrated Multiplexed Readout}
\author{M.S. Allman}
\author{V.B. Verma}
\author{M. Stevens}
\author{T. Gerrits}
\author{R.D. Horansky}
\author{A.E. Lita}
\affiliation{National Institute of Standards and Technology, 325 Broadway, Boulder, Colorado 80305-3328, USA}
\author{F. Marsili}
\author{A. Beyer}
\author{M.D. Shaw}
\affiliation{Jet Propulsion Laboratory, 4800 Oak Grove Dr., Pasadena, California 91109, USA}
\author{D. Kumor}
\affiliation{Purdue University, 610 Purdue Mall, West Lafayette, Indiana 47907, USA}
\author{R. Mirin}
\author{S.W. Nam}
\affiliation{National Institute of Standards and Technology, 325 Broadway, Boulder, Colorado 80305-3328, USA}
\email{shane.allman@boulder.nist.gov}
\date{\today}

\begin{abstract}
We demonstrate a 64-pixel free-space-coupled array of superconducting nanowire single photon detectors optimized for high detection efficiency in the near-infrared range.  An integrated, readily scalable, multiplexed readout scheme is employed to reduce the number of readout lines to 16.  The cryogenic, optical, and electronic packaging to read out the array, as well as characterization measurements are discussed.
\end{abstract}
\maketitle

Superconducting nanowire single photon detectors (SNSPD) have been shown to have high efficiency, low dark counts, and tens of picosecond timing \cite{Goltsman2001}.  SNSPDs have been particularly useful in applications requiring high timing resolution and detection in the near-infrared ($\lambda > 1\mu$m)\cite{Natarajan2012}.  Until recently small arrays of nanowire detectors for imaging, higher count rates, large collection areas, and photon number resolving detection have been technologically challenging to realize.  The recent observation of the saturation of internal detection efficiency at $\sim40$\% of the maximum bias current in SNSPDs fabricated from amorphous tungsten silicide (WSi) is key in enabling high-efficiency arrays to be constructed \cite{Baek2011,Marsili2013}.  Detectors fabricated from niobium nitride (NbN), for example, have high detection efficiencies only at bias currents close to the critical current \cite{Rosenberg2013}.  Current ``cross talk'' between detectors in arrays biased so close to their maximum operating current could cause other detectors to falsely fire when one detector in the array fires.  A wide margin in operating bias, or ``bias plateau'', allows the detectors to be biased at a fraction of their respective critical currents and to remain sensitive to photons, even as other detectors in the array fire.  Previously, we demonstrated a $4$-pixel WSi SNSPD array with an integrated, scalable multiplexed readout \cite{Verma2014}.  In this work we extend to a free-space coupled $64$-pixel ($8\times8$ square) array using a slightly modified version of our multiplexed readout.

To date, only a handful of experiments have demonstrated arrays of SNSPDs.  Architectures where each detector has its own readout/bias line have been measured \cite{Divochiy2008, Rosenberg2013, Miki2014, Shaw2014}.  However, one critical issue to consider when scaling up to larger numbers of elements is the available cooling power of the cryogenic system.  Each readout channel adds to the total thermal power budget and can quickly limit achievable base temperatures.  Therefore, multiplexing schemes, where the total number of readout channels is kept to a small fraction of the number of array elements, become a necessity.  Attempts at multiplexed readout have primarily been limited to single flux quantum (SFQ) logic schemes \cite{Hofherr2012, Ortlepp2011, Miki2011, Terai2012, Yamashita2012}.  SFQ-logic-based readout is attractive due to the intrinsic low power consumption but the designs can be quite complex and require additional fabrication steps for the Josephson junctions.  Apart from SFQ, an inductive current splitting technique where the firing pixel location was encoded onto the magnitude of the output pulse, has been demonstrated in a $4$-pixel linear array \cite{Zhao2013}.

Another key issue in array development is device yield.  The traditional materials used to date for making SNPSD arrays has been limited to the NbN family of materials and related alloys.  Achieving high detection efficiencies with this material however requires wire widths $<100$ nm which, from a fabrication standpoint, can prove difficult in achieving the necessary uniformity among all the pixels \cite{Gaudio2014}.  The result is that single-pixel active areas are relatively small compared to WSi, making efficient optical coupling more difficult, particularly in a free-space-coupled setup.  With WSi, saturation in detection efficiency can be achieved with wire widths $>100$ nm, which, in turn, allows devices with larger active areas to be fabricated with higher yield.  To the best of our knowledge, there have been no experiments demonstrating arrays larger than $12$ pixels and combining a fully-integrated, multiplexed, readily scalable, readout with both temporal and spatial resolution of photo-detection events in a single free-space-coupled experiment.

\begin{figure*}[!htbp]
\centering
\includegraphics[width=2\columnwidth]{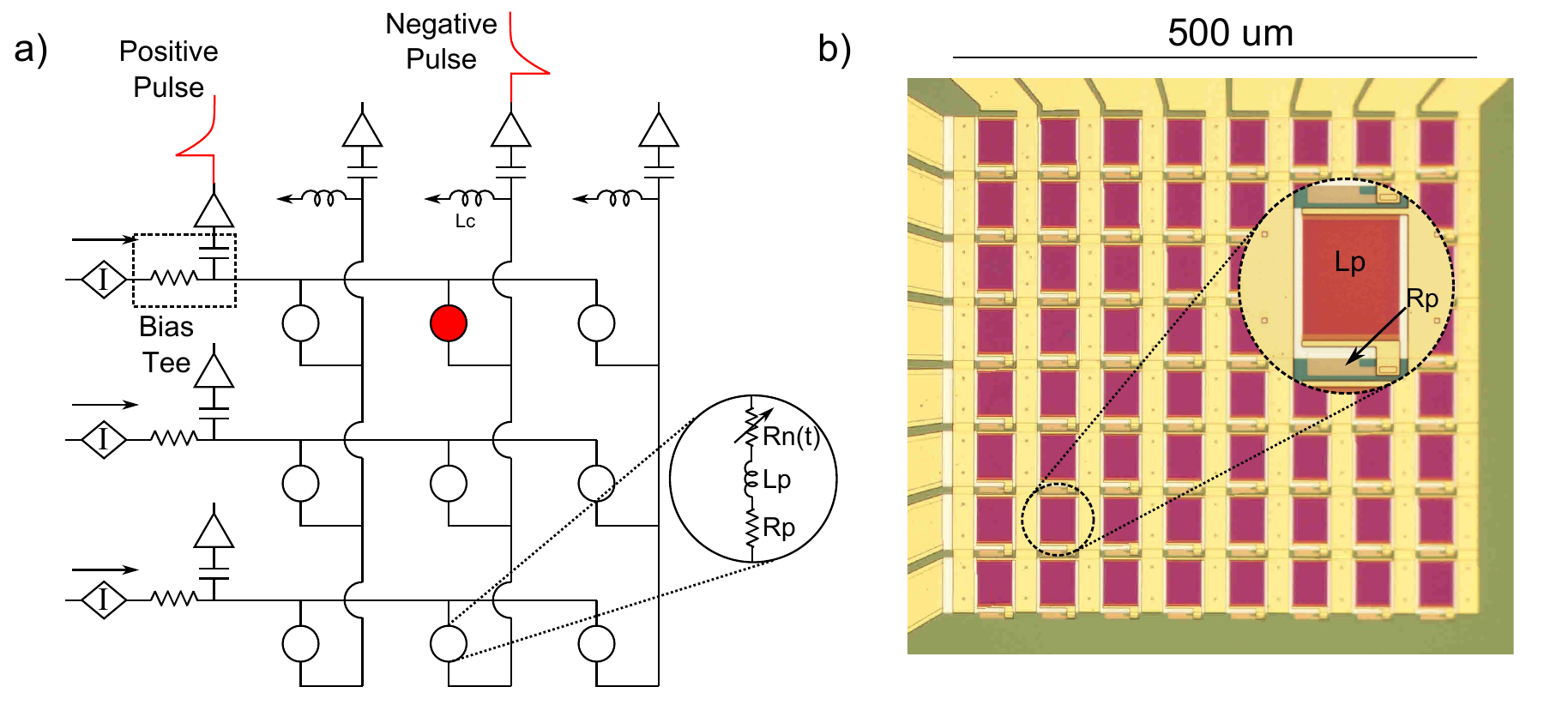}
\caption{a) A $3\times3$ array illustrating the multiplexed readout. Bias current is introduced to each row and shunted to ground by $L_c$.  When a given pixel fires, current is diverted out of the pixel resulting in a positive voltage pulse from the row amplifiers and a negative pulse in a column amplifier.  Comparison between pulse arrival times between the row and column amplifiers reveal which pulse fired. b) Optical micrograph of the array.}
\label{RCReadout}
\end{figure*}
A schematic illustrating our approach to building a detector array is shown in Fig \ref{RCReadout}.  As with our $2\times2$ demonstration experiment, the readout is multiplexed, requiring only $2N$ readout channels for an $N\times N$ square array (Figure 1 shows a $3\times3$ for simplicity).  All N detectors in a given row are wired in parallel to a common current bias introduced through the ``dc'' port of a resistive bias-T.  Additionally, an ac-coupled row readout amplifier is connected to the ``rf'' port of the bias-T.  Each detector consists of an SNSPD with kinetic inductance $L_p\sim8 \mu$H in series with a resistance $R_p\sim50\Omega$ that ensures the bias current is partitioned equally among each detector. The row current path to ground is provided by an inductance $L_c\sim3 \mu$H.  When a given detector in the row absorbs a photon, a local ``hot spot'' is created, causing that region of the SNSPD to switch to the normal state with resistance $R_N$ \cite{Yang2007}.  Current is then diverted out of that detector (on a time scale $\tau_R\sim L_p/R_N\sim1$ ns) and into the row readout amplifier where a positive voltage pulse is observed.  This part of the readout, however, only identifies the row in which a photon was absorbed.  To identify which detector in a given row absorbed a photon, an ac-coupled column readout amplifier is wired in parallel with $L_c$ at the base of each pixel.  Over the time $\tau_R$, the impedance of the column inductor is much larger than the column amplifier impedance ($L_c/\tau_R>>R_a=50\Omega$).  In our previous experiment, the row current path to ground was provided by a resistance (labeled $R_R$ in the previous paper).  As a result, the amplitude of the column pulse was reduced by the factor $R_R/\left(R_R+R_a\right)$.  Increasing $R_R$ would improve the signal amplitude but at the expense of increased power dissipation on chip.  Our solution was to replace $R_R$ with $L_c$.  As such, current is preferentially diverted from the column readout amplifier resulting in a negative pulse whose amplitude is very close in magnitude the amplitude of the row pulse.  The pulses are fed to room-temperature amplifiers and comparator circuits and then time-stamped using a $16$-channel time-tagging unit.  Comparison of pulse arrival times from the positive pulse from one of the N row amplifiers along with the negative pulse from one of the N column amplifiers reveals which of the $N^2$ pixels in the array fired.

An optical micrograph of the array is shown in Fig 1 b.  The array was fabricated on a $3$" silicon (Si) wafer. Silver leads for the row readout lines and ground plane were deposited by electron beam evaporation and patterned by liftoff.  The wiring layer also served as a mirror under each SNSPD.  Silver was chosen to reduce stray resistance in the wiring lines that could adversely affect the current partitioning among the detectors, and to reduce roughness of the silicon dioxide (SiO$2$) surface described below.  The series resistance $R_p$ was fabricated using palladium gold (PdAu). The main device layer consists of a $4.5$ nm-thick layer of WSi deposited by DC magnetron co-sputtering from separate W and Si targets.  A $235$ nm-thick layer of SiO$2$, serves as the inter-layer dielectric with $2\mu$m square vias etched though the SiO$2$ to make the necessary wiring connections between the device and bias lines.  Electron beam lithography was then performed to define the nanowires in PMMA.  Each pixel has an area of $30\mu$m $\times$ $30\mu$m, and consists of meandering $160$nm-wide nanowires on a $240$nm pitch.  The pixels are spaced on a $60\mu$m pitch.  The $160$ nm nanowire widths were chosen to maximize yield at the expense of a large plateau in detection efficiency vs. current bias.  Typical single-pixel devices are fabricated with $120-160$ nm-wide nanowires \cite{Marsili2013, Verma2014}.

\begin{figure}[!htbp]
\centering
\includegraphics[width=0.8\columnwidth]{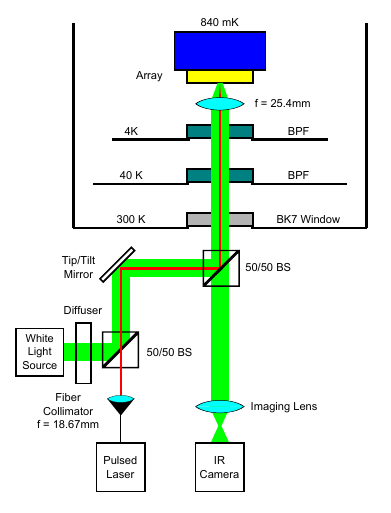}
\caption{Free-space-coupling diagram.  Collimated light from a pulsed laser, as well as incoherent white light, was introduced into the cryostat through two $50/50$ beam splitters.  The cryostat window was $0.5$ inches in diameter and made from $5$mm thick BK7 glass.  A $40$nm wide BPF, centered at $1550$nm and mounted to the $40$K stage and a $500$nm wide BPF, centered at $1750$nm and mounted to the $4$K stage, was used to filter black body radiation.  An AR-coated lens with a $25.4$mm focal length and $1$inch diameter was positioned directly in front of the array to focus the collimated beam to a small spot size.  Light scattered off the array was sent to an IR camera to aid in alignment of the beam to the array.}
\label{FSSetup}
\end{figure}
The array was cooled to $\sim840$mK using a sealed helium-$4$ adsorption refrigerator mounted to a pulse tube cooler.  The refrigerator had a hold time of $\sim40$hrs with the readout lines and optical windows installed.  The readout lines in the cryostat consisted of $16$ SMA coaxial cables with appropriately chosen thermal conductance to keep the heat load on the $800$mK stage to $\sim50\mu$W.  Light was free-space-coupled to the array from outside the cryostat as shown in Fig \ref{FSSetup}.  $1550$nm light from a $50$MHz mode-locked laser source was fiber-coupled using single-mode fiber with a $10 \mu$m core diameter and $0.1$ numerical aperture to a fiber collimation package with an $18.67$mm focal length.  The collimated beam was then sent into the cryostat through two $50/50$ beam splitters.  Alignment of the beam to the array was controlled using a single tip/tilt mirror.  Incoherent light from an incandescent source, used to flood illuminate the array, was sent through a diffuser and into the cryostat through the same two beam splitters.  Light scattered off the array was imaged with an infrared camera outside the cryostat to aid in alignment of the beam to the array.  Light entered the cryostat through a $0.5$ inch diameter window made from $5$mm thick BK7.  From here the light passed through a series of band pass filters (BPF) to filter out room temperature black body radiation.  The first BPF on the $40$K stage had a $40$nm bandwidth centered at $1550$nm.  The second BPF on the $4$K stage had a $500$nm bandwidth centered at $1750$nm.  The filter combination was selected to give good rejection of light at wavelengths $>2\mu$m using commercially-available components.

\begin{figure}[!htbp]
\centering
\includegraphics[width=1\columnwidth]{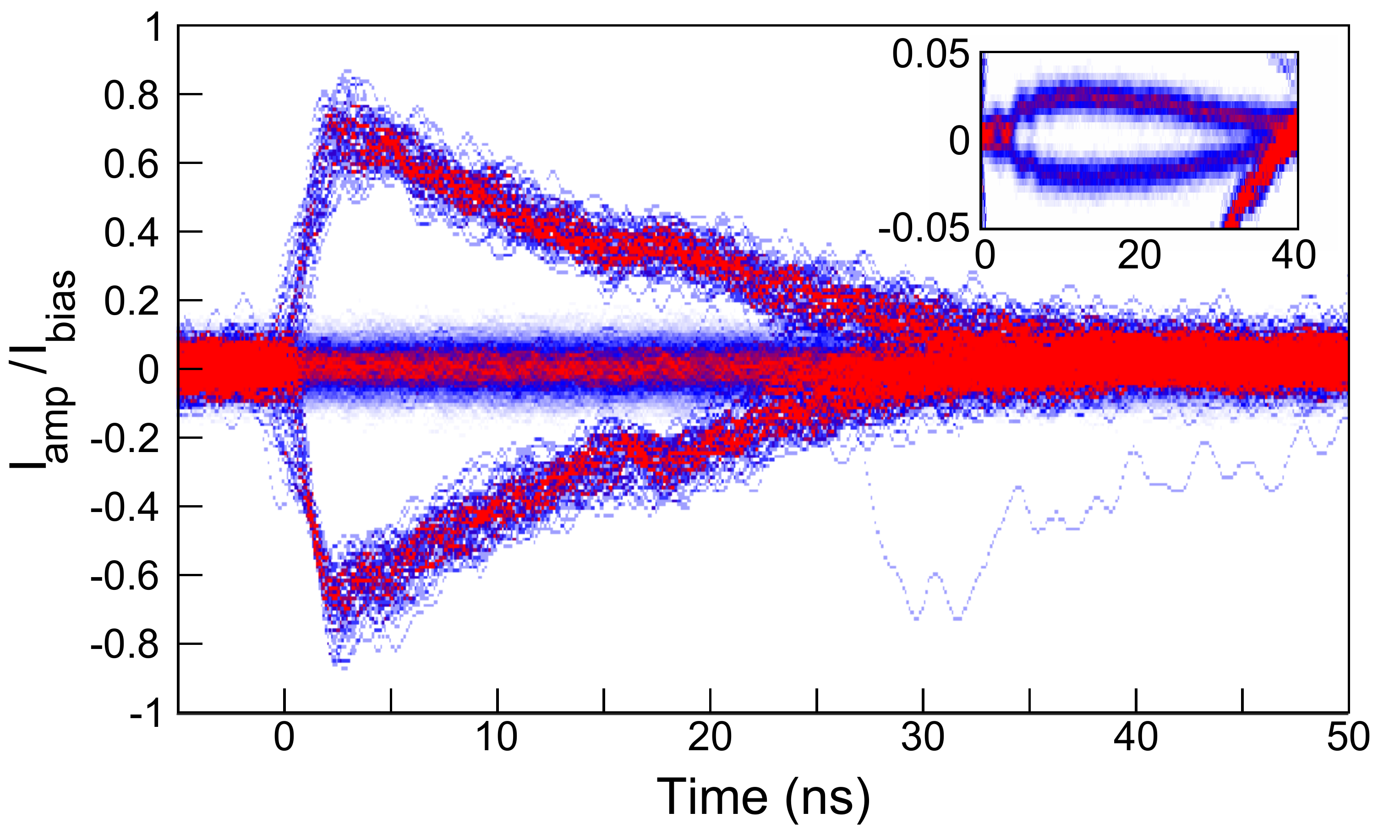}
\caption{Persistence plot of the row and column current pulses as well as the electrical cross talk in non-firing channels.  The amplitudes were normalized to the per-pixel applied bias current.  The inset shows averaged electrical cross talk pulses that were well below the system noise level in the main figure.}
\label{Pulses}
\end{figure}
In Fig \ref{Pulses} we plot single-shot traces of the readout pulses for all $64$ pixels in the array.  The traces were captured using a $16$ channel oscilloscope by biasing each row individually and separately triggering on each of the $8$ column readout channels.  When a row and column pulse were recorded, the signals from the remaining row and column channels were also captured in order to show the level of electrical cross-talk between readout channels.  The large-amplitude positive pulses were from the row readout amplifiers and the negatively large amplitude pulses were from the column readout amplifiers.  The electrical cross-talk was smaller than the system noise level and could not be resolved in the main figure.  In the inset, we resolve the cross-talk traces by averaging $50$ times.  We can see that the column pulse amplitudes were nearly equal to the row pulse amplitudes.  This equality was due to the fact that the column inductor's impedance was much larger than the column amplifier impedance and marked an improvement over the readout design employed in the $2\times2$ experiment.  The pulse amplitudes were calculated using the measured gain of the amplifiers ($59$dB) and assuming $50\Omega$ amplifier input impedances.  The amplitudes were then normalized by the per-pixel applied bias current which was assumed to be uniform.  We note that the current pulse amplitudes coupled into the amplifiers were only about $80\%$ of the current in each pixel (a small port reflection was present at the input of the amplifiers).  The remaining $20\%$ of the current was distributed throughout the array in the form of leakage current to non-firing pixels.  We have developed a model, discussed in the supplemental material, that allows us to quantify the amplitude dependence on array size.  As the number of array elements is increased, the pulse amplitudes are expected to further decrease using this type of multiplexing.  Once the pulse heights become comparable to the system rms noise level, discrimination between the pulses and system noise will be difficult to achieve without advanced signal processing techniques. Our model predicts that the maximum array size we can achieve is $225\times225$, with identical electrical parameters used in this experiment.  The details of this calculation are discussed in the supplemental material \cite{Supplemental}.

\begin{figure*}[!htbp]
\centering
\includegraphics[width=2\columnwidth]{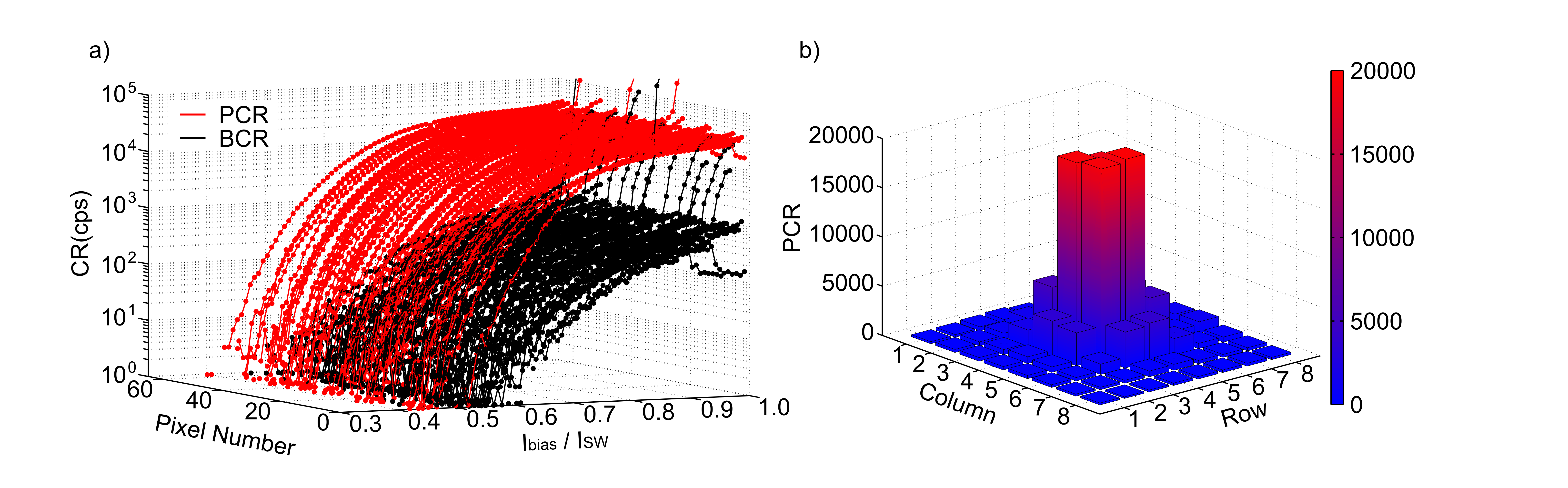}
\caption{a) PCR and BCR for all $64$ pixels as a function of bias current applied to all $8$ rows simultaneously. b) Beam image with the rows biased at $I_{bias} = 0.92I_{SW}$ showing the spatial resolution capabilities of the array and readout.}
\label{CRvsBias}
\end{figure*}
For the first demonstration of the array, we characterized the count rate (CR) of the array as a function of bias current applied to each row simultaneously.  The background count rate (BCR) was measured with all sources of light turned off.  The photo-count rate (PCR) was measured with the array flood illuminated using the incandescent source.  All $16$ channels were read out simultaneously using a $16$ channel time-tagging unit.  For each bias current value, the count rate was integrated over $1$ second.  Figure \ref{CRvsBias} a) shows a plot of the counts recorded for all $64$ pixels in the array.  The red traces are the per-pixel PCR and the black traces are the per-pixel BCR.  The applied bias current for each trace was normalized by the maximum operating current or ``switching'' current $I_{SW}$ for that row.  For each row, $I_{SW}\sim28.5$ $\mu$A when not illuminated and $I_{SW}\sim27.3$ $\mu$A when illuminated with the incandescent source.  We attribute the reduction in $I_{switch}$ to a count-rate-dependent re-biasing effect due to the AC coupling of the readout channels to the readout amplifiers \cite{Kerman2013}.  The observed BCR was $\sim1$ kHz per pixel, and for the flood-illumination intensity used in the figure, the observed PCR was $\sim30$kHz per pixel at $I_{bias}\sim I_{SW}$.  We note that the plateau in PCR with bias current wasn't as large as we typically observe in single-pixel devices.  Additional work needs to be done to understand why.  In Figure \ref{CRvsBias} b) we show the image of the collimated beam from the mode-locked laser with each row biased at $I_{bias} = 0.92 I_{switch}$.  A long focal length lens outside the cryostat was used to adjust the size of the beam to overlap with $\sim16$ pixels in the center of the array.  The spatial distribution of the counts in the image is consistent with what we expect to see from a Gaussian beam and indicates that the row/column readout provides the spatial resolution we expect.

In summary, we have demonstrated a free-space-coupled $64$ pixel WSi SNSPD array using an integrated, scalable multiplexed readout architecture.  Each pixel has been shown to be sensitive to light and the row/column multiplexing scheme is capable of spatially resolving detection events.  We have improved the signal to noise ratio for the column readout pulses by replacing the column resistance with an inductance.  We characterized the array using free-space-coupled light from an incoherent source as well as coherent light from a laser.  This work was supported by NIST and the DARPA INPHO program.
%
%

\end{document}